\documentclass[twocolumn,showpacs,preprintnumbers,amsmath,amssymb]{revtex4}

\usepackage{graphicx}
\usepackage{dcolumn}
\usepackage{bm}

\newcommand{\bq}{\begin{equation}}
\newcommand{\eq}{\end{equation}}
\newcommand{\bqa}{\begin{eqnarray}}
\newcommand{\eqa}{\end{eqnarray}}
\newcommand{\nn}{\nonumber \\}
\newcommand{\ij}{\langle i j \rangle}

\def\be     {\begin{equation}}
\def\ee     {\end{equation}}
\def\bea        {\begin{eqnarray}}
\def\eea        {\end{eqnarray}}
\def\bnn    {\begin{eqnarray*}}
\def\enn    {\end{eqnarray*}}

\begin{document}

\title{Competition between superconductivity and charge density waves}
\author{Ki-Seok Kim}
\affiliation{School of Physics, Korea Institute for Advanced
Study, Seoul 130-012, Korea}
\date{\today}

\begin{abstract}
We derive an effective field theory for the competition between
superconductivity (SC) and charge density waves (CDWs) by
employing the SO(3) pseudospin representation of the SC and CDW
order parameters. One important feature in the effective nonlinear
$\sigma$ model is the emergence of Berry phase even at half
filling, originating from the competition between SC and CDWs,
i.e., the pseudospin symmetry. A well known conflict between the
previous studies of Oshikawa\cite{Oshikawa} and D. H. Lee et
al.\cite{DHLee} is resolved by the appearance of Berry phase. The
Berry phase contribution allows a deconfined quantum critical
point of fractionalized charge excitations with $e$ instead of
$2e$ in the SC-CDW quantum transition at half filling.
Furthermore, we investigate the stability of the deconfined
quantum criticality against quenched randomness by performing a
renormalization group analysis of an effective vortex action. We
argue that although randomness results in a weak disorder fixed
point differing from the original deconfined quantum critical
point, deconfinement of the fractionalized charge excitations
still survives at the disorder fixed point owing to a nonzero
fixed point value of a vortex charge.
\end{abstract}


\maketitle

\section{Introduction}

Recently, it was proposed that when there exist two competing
orders characterized by different patterns of symmetry breaking,
the two order parameters can acquire some topological Berry phases
to allow a continuous quantum phase transition between the two
states, although forbidden in the Landau-Ginzburg-Wilson (LGW)
theoretical framework without fine-tuning of couplings admitting
multi-critical points.\cite{Tanaka,Senthil_superspin} Especially,
the quantum critical point in this quantum phase transition is
quite exotic in the respect that elementary excitations are
fractionalized, thus called a deconfined quantum critical
point.\cite{Senthil_DQCP,Kim_DQCP}

One deconfined quantum critical point was demonstrated in the
competition between antiferromagnetic (AF) and valance bond solid
(VBS) orders.\cite{Senthil_DQCP,Kim_DQCP} Tanaka and Hu considered
an SO(5) superspin representation including both the AF and VBS
order parameters, and derived an effective nonlinear $\sigma$
model for the SO(5) superspin variable from the spinon
representation of the Heisenberg Hamiltonian.\cite{Tanaka} One
crucial feature in their effective field theory is the presence of
Berry phase for the superspin field. They demonstrated that the
competition between AF and VBS is well described by the SO(5)
nonlinear $\sigma$ model with a topological Berry phase term.

In the present paper we consider another concrete example, the
competition between superconductivity (SC) and charge density
waves (CDWs), as a simplified version of the competition between
AF and VBS. Introducing an SO(3) pseudospin representation to
include both the SC and CDW order parameters, we derive an
effective nonlinear $\sigma$ model in terms of the O(3) pseudospin
variable from the attractive Hubbard model. Interestingly, a Berry
phase term naturally appears in this $\sigma$ model, allowing a
deconfined quantum critical point of fractionalized charge
excitations with $e$ instead of $2e$ as a result of the
competition between SC and CDW. Furthermore, we examine the
stability of the deconfined quantum criticality against quenched
randomness generating two kinds of random potentials, a random
mass term and a random fugacity one in the effective vortex action
[Eq. (16)]. Performing a renormalization group (RG) analysis of
the vortex action [Eq. (16)] in the London approximation [Eq.
(17)], we argue that deconfinement of the fractionalized
excitations still survives although the presence of disorder leads
to a new quantum critical point with finite disorder strength. We
find that the stability of the deconfined quantum criticality
originates from the existence of the charged critical point.

Before going further, it is valuable to address several important
differences between the present work and previous studies. Earlier
studies\cite{Old_SC_CDW} revealed that the half-filled negative-U
Hubbard model on a 2d square lattice is mathematically equivalent
to the positive-U Hubbard model, using the particle-hole
transformation. This equivalence maps the XY ordered
antiferromagnetic phase of the spin system that results for
positive-U to the superfluid phase of the negative-U problem.
Likewise, the Ising antiferromagnet (for positive U) maps to a CDW
phase (for negative U). However, in these earlier
studies\cite{Old_SC_CDW} the role of Berry phase was not
investigated clearly, thus the LGW-forbidden continuous transition
and deconfined quantum critical points were not found in the
context of SC-CDW transitions.

It is interesting to understand the origin of the Berry phase in
the negative-U Hubbard model and the positive-U one. The
positive-U Hubbard model reduces to the antiferromagnetic
Heisenberg model in the large-U limit. In the negative-U Hubbard
model the low energy effective action can be mapped onto an
effective model of hard-core lattice bosons with a hopping
amplitude of order $t^2/U$ and repulsive nearest neighbor
interaction of the same order in the strong coupling limit $U
\rightarrow -\infty$.\cite{Hard_core_boson} One can show that this
hard-core boson model is equivalent to the antiferromagnetic
Heisenberg model, associated with charge degrees of freedom to
form a pseudospin.\cite{Old_SC_CDW,Zhang} The Berry phase in the
negative-U Hubbard model originates from the pseudospin (charge)
SU(2) symmetry\cite{Old_SC_CDW} while it in the positive-U Hubbard
model comes from the spin SU(2) symmetry. It should be noted that
this topological phase appears even at half filling. On the other
hand, it was not allowed at half filling in recent
studies.\cite{Balents,Tesanovic} The Berry phase resulting from
the chemical potential in the boson Hubbard-type
model\cite{Balents,Tesanovic} is different from the present one
because the presence of the chemical potential reduces the SU(2)
pseudospin symmetry to the U(1) one. This is the reason why there
exists only the Berry phase coming from the chemical potential in
the boson Hubbard-type model while our effective action has both
Berry phases resulting from the SU(2) pseudospin symmetry and
chemical potential. In other words, the competition between SC and
CDWs results in a non-trivial Berry phase term even at half
filling. Thus, the chemical potential plays the role of an
additional Berry phase in the present effective theory.
Furthermore, the appearance of Berry phase at half filling allows
other possible disordered phases corresponding to valance bond
orders in the pseudospin language. This resolves the well known
conflict between the two previous studies\cite{Oshikawa,DHLee}
that Ref. \cite{DHLee} does not admit a dimerised order while the
paper \cite{Oshikawa} claims this phase is certainly possible. The
emergence of Berry phase at half filling clearly reveals how the
dimerised order appears.

We would like to mention that the present quantum transition
occurs between the XY ordered phase and the Ising
antiferromagnetic one if one maps our negative-U problem to the
positive-U one. This XY-Ising antiferromagnetic transition allows
the SO(3) pseudospin description for the competition of SC and CDW
fluctuations in the context of the negative-U Hubbard model. On
the other hand, the AF-VBS quantum transition requires the SO(5)
superspin description for the competition of AF and VBS
fluctuations.\cite{Tanaka}

\section{Effective field theory}

\subsection{Derivation of the O(3) nonlinear $\sigma$ model from the attractive Hubbard model}

We consider the attractive Hubbard Hamiltonian \bqa && H = -
t\sum_{ij\sigma}c_{i\sigma}^{\dagger}e^{iA_{ij}}c_{j\sigma} -
\frac{3u}{2}\sum_{i}c_{i\uparrow}^{\dagger}c_{i\uparrow}
c_{i\downarrow}^{\dagger}c_{i\downarrow} \nn && -
\sum_{i\sigma}v_{i}c_{i\sigma}^{\dagger}c_{i\sigma}. \eqa Here $t$
is a hopping integral of electrons, and $u$ strength of on-site
Coulomb repulsions. $A_{ij}$ is an external (static)
electromagnetic field, and $v_{i}$ a quenched random potential.

The local interaction term can be decomposed into pairing and
density channels in the following way \bqa &&
-\frac{3u}{2}\sum_{i}c_{i\uparrow}^{\dagger}c_{i\uparrow}c_{i\downarrow}^{\dagger}c_{i\downarrow}
= -
\frac{u}{2}\sum_{i}c_{i\uparrow}^{\dagger}c_{i\downarrow}^{\dagger}
c_{i\downarrow}c_{i\uparrow} \nn && -
\frac{u}{2}\sum_{i}\Bigl(\sum_{\sigma}c_{i\sigma}^{\dagger}c_{i\sigma}-1\Bigr)^{2}
-
\frac{u}{2}\Bigl(\sum_{\sigma}c_{i\sigma}^{\dagger}c_{i\sigma}-1\Bigr)
. \nonumber \eqa Performing the Hubbard-Stratonovich
transformation for the pairing and density interaction channels,
we find an effective Lagrangian in the Nambu-spinor representation
\bqa && Z = \int{D[\psi_{i},\psi_{i}^{\dagger}, \Phi^{R}_{i},
\Phi_{i}^{I}, \varphi_{i}]}e^{-\int{d\tau} L} , \nn && L =
\sum_{i}\psi_{i}^{\dagger}(\partial_{\tau}\mathbf{I} -
\mu\tau_{3})\psi_{i} -
t\sum_{\ij}(\psi_{i}^{\dagger}\tau_{3}e^{iA_{ij}\tau_{3}}\psi_{j}
+ H.c.) \nn && -
\sum_{i}(\Phi^{R}_{i}\psi_{i}^{\dagger}\tau_{1}\psi_{i} +
\Phi^{I}_{i}\psi_{i}^{\dagger}\tau_{2}\psi_{i} +
\varphi_{i}\psi_{i}^{\dagger}\tau_{3}\psi_{i}) \nn && +
\frac{1}{2u}\sum_{i}(\Phi^{R2}_{i} + \Phi^{I2}_{i} +
\varphi_{i}^{2}) -
\sum_{i}v_{i}(\psi_{i}^{\dagger}\tau_{3}\psi_{i}+1) . \eqa Here
$\psi_{i}$ is the Nambu spinor, given by
$\psi_{i}= \left(\begin{array}{c} c_{i\uparrow} \\
c_{i\downarrow}^{\dagger} \end{array} \right)$.  $\Phi^{R}_{i}$
and $\Phi^{I}_{i}$ are the real and imaginary parts of the
superconducting order parameter respectively, and $\varphi_{i}$ an
effective density potential. $\mu$ is an electron chemical
potential which differs from its bare value $\mu_{b}$ as $\mu =
\mu_{b} + u/2$.

Introducing a pseudospin vector $\vec{\Omega}_{i} \equiv
(\Phi_{i}^{R}, \Phi_{i}^{I}, \varphi_{i})$, one can express Eq.
(2) in a compact form \bqa && Z =
\int{D[\psi_{i},\psi_{i}^{\dagger},
\vec{\Omega}_{i}]}e^{-\int{d\tau} L} , \nn && L =
\sum_{i}\psi_{i}^{\dagger}\partial_{\tau}\psi_{i} -
t\sum_{\ij}(\psi_{i}^{\dagger}\tau_{3}e^{iA_{ij}\tau_{3}}\psi_{j}
+ H.c.) \nn && -
\sum_{i}\psi_{i}^{\dagger}(\vec{\Omega}_{i}\cdot\vec{\tau})\psi_{i}
+
\frac{1}{4u}\sum_{i}\mathbf{tr}[\vec{\Omega}_{i}\cdot\vec{\tau}-(\mu+v_{i})\tau_{3}]^{2}
\nn && - \sum_{i}v_{i} , \eqa where we used the shift of
$\varphi_{i} \rightarrow \varphi_{i} - \mu - v_{i}$. Integrating
over the pseudospin field $\vec{\Omega}_{i}$, Eq. (3) recovers the
Hubbard model Eq. (1).

In this paper we consider only phase fluctuations in ${\vec
\Omega}_{i}$, assuming amplitude fluctuations frozen thus setting
it as ${\vec \Omega}_{i} = m{\vec n}_{i}$ with an amplitude $m$.
Since our starting point is a nonzero amplitude of the pseudospin
field, we utilize a strong coupling approach decomposing the
directional fluctuating field ${\vec n}_{i}$ into two complex
boson fields, so called $CP^{1}$ representation\cite{Kim_Kondo}
\bqa && {\vec n}_{i}\cdot{\vec \tau} =
U_{i}\tau^{3}U_{i}^{\dagger} , \nn && U_{i} = \left( \begin{array}{cc} z_{\uparrow} & -z_{\downarrow}^{\dagger} \\
z_{\downarrow} & z_{\uparrow}^{\dagger} \end{array} \right) , \eqa
where $U_{i}$ is an SU(2) matrix field in terms of a complex boson
field $z_{i\sigma}$ with pseudospin $\sigma$. Using the $CP^{1}$
representation in Eq. (3), and performing the gauge transformation
\bqa && \Psi_{i} = U^{\dagger}_{i}\psi_{i} , \eqa Eq. (3) reads
\bqa && Z =
\int{D[\Psi_{i},\Psi_{i}^{\dagger},U_{i}]}e^{-\int{d\tau} L} , \nn
&& L = \sum_{i}\Psi_{i}^{\dagger}(\partial_{\tau}\mathbf{I} -
m\tau_{3} + U_{i}^{\dagger}\partial_{\tau}U_{i})\Psi_{i} \nn && -
t\sum_{\ij}(\Psi_{i}^{\dagger}U_{i}^{\dagger}\tau_{3}e^{iA_{ij}\tau_{3}}U_{j}\Psi_{j}
+ H.c.) \nn && + \frac{1}{4u}\sum_{i}\mathbf{tr}[m\tau_{3} -
(\mu+v_{i}){U}_{i}^{\dagger}\tau_{3}U_{i}]^{2} - \sum_{i}v_{i} .
\eqa

Since Eq. (6) is quadratic for the spinor field $\Psi_{i}$, one
can formally integrate out the spinor field to obtain \bqa &&
S_{eff} = - \mathbf{tr}ln\Bigl[\partial_{\tau}\mathbf{I} -
m\tau_{3} + U_{i}^{\dagger}\partial_{\tau}U_{i} -
t_{ij}U_{i}^{\dagger}\tau_{3}e^{iA_{ij}\tau_{3}}U_{j}\Bigr] \nn &&
+ \int{d\tau}\Bigl[ - \frac{m}{2u}\sum_{i}(\mu +
v_{i})\mathbf{tr}[{U}_{i}^{\dagger}\tau_{3}U_{i}\tau_{3}] \nn && +
\sum_{i}\Bigl(\frac{v_{i}^{2} + \mu^{2} + m^2 + \mu{v}_{i}}{2u} -
v_{i} \Bigr)\Bigr] . \eqa Expanding the logarithmic term for
$U_{i}^{\dagger}\partial_{\tau}U_{i}$ and
$U_{i}^{\dagger}\tau_{3}e^{iA_{ij}\tau_{3}}U_{j}$, we obtain \bqa
&& S_{eff} \approx
\sum_{i}\mathbf{tr}[G_{0}(U_{i}^{\dagger}\partial_{\tau}U_{i})]
\nn && +
\frac{1}{2}\sum_{i}\mathbf{tr}_{j}[G_{0}t_{ij}U_{i}^{\dagger}\tau_{3}
e^{iA_{ij}\tau_{3}}U_{j}G_{0}t_{ji}U_{j}^{\dagger}\tau_{3}e^{-iA_{ij}\tau_{3}}U_{i}]
\nn && + \int{d\tau}\Bigl[ - \frac{m}{2u}\sum_{i}(\mu +
v_{i})\mathbf{tr}[{U}_{i}^{\dagger}\tau_{3}U_{i}\tau_{3}] \nn && +
\sum_{i}\Bigl(\frac{v_{i}^{2} + \mu^{2} + m^2 + \mu{v}_{i}}{2u} -
v_{i} \Bigr)\Bigr] , \eqa where $G_{0} = -
(\partial_{\tau}\mathbf{I} - m\tau_{3})^{-1}$ is the single
particle propagator. The first term leads to Berry phase while the
second results in an exchange interaction term. The resulting
effective action is obtained to be without the electromagnetic
field $A_{ij}$ \bqa && S_{eff} =
iS\sum_{i}\omega(\{\mathbf{S}_{i}(\tau)\}) +
\int_{0}^{\beta}{d\tau} H_{eff} , \nn && H_{eff} = -
J\sum_{ij}(S_{i}^{x}S_{j}^{x} + S_{i}^{y}S_{j}^{y}) +
V\sum_{ij}S_{i}^{z}S_{j}^{z} \nn && - \sum_{i}(\mu +
v_{i})S_{i}^{z} , \eqa where the effective exchange coupling
strength is given by $J = V = 2t^2/m$.\cite{Wen,Nagaosa} It is
interesting that the effective Hamiltonian for the competition
between SC and CDW is obtained to be the Heisenberg model in terms
of the O(3) pseudospin variable. One important message in this
effective action is that the Berry phase term
$iS\sum_{i}\omega(\{\mathbf{S}_{i}(\tau)\})$ should be taken into
account for the SC-CDW transition even at half filling.
Furthermore, the chemical potential plays the same role as an
external magnetic field, and the disorder potential a random
magnetic field.

If we consider half filling without disorder, i.e., $\mu = v_{i} =
0$, the XY order of $\langle{S}_{i}^{\pm}\rangle \not= 0$ and
$\langle{S}_{i}^{z}\rangle = 0$ is expected in the case of $J
>> V$, identified with SC. On the other hand, the Ising order of
$\langle{S}_{i}^{z}\rangle \not= 0$ and
$\langle{S}_{i}^{\pm}\rangle = 0$ arises in the case of $V >> J$,
corresponding to CDW because of the Berry phase, as will be
discussed below. One important question in this paper is how the
SC-CDW transition appears in the presence of disorder.

It is easy to show that the Heisenberg model with ferromagnetic XY
couplings is the same as that with antiferromagnetic ones.
Performing the Haldane mapping of the antiferromagnetic Heisenberg
model\cite{Nagaosa} with a magnetic field in the $z$-direction, we
obtain the O(3) nonlinear $\sigma$ model \bqa && S_{\sigma} =
iS\sum_{i}(-1)^{i}\omega(\{\mathbf{n}_{i}(\tau)\}) +
\frac{1}{g}\int_{0}^{c\beta}dx_{0}\int{d^d\mathbf{x}}\Bigl[(\partial_{0}n_{z})^{2}
\nn && + (\partial_{0}n_{x} - i[\mu+v]n_{y})^{2} +
(\partial_{0}n_{y} + i[\mu+v]n_{x})^{2} +
(\nabla_{\mathbf{x}}\mathbf{n})^{2} \Bigr] , \nn \eqa where $c$ is
the velocity of spin waves, and $g$ the coupling strength between
spin wave excitations. As Tanaka and Hu derived an effective SO(5)
nonlinear $\sigma$ action of the superspin field for the AF-VBS
transition, we derived an effective SO(3) nonlinear $\sigma$
action of the pseudospin field for the SC-CDW transition.
Furthermore, this effective $\sigma$ action includes not only
doping contributions but also disorder effects. On the other hand,
in the SO(5) superspin $\sigma$ model it is not clear how the
doping effect modifies the effective action because a chemical
potential term breaks the relativistic invariance. In this case it
is not clear even to obtain the topological term. In the following
we discuss how this $\sigma$ action describes the competition
between SC and CDW in the presence of quenched disorder by
focusing on the role of Berry phase.

Without loss of generality we use the parametrization \bqa &&
{\vec n}_{i} = (\sin(u\vartheta_{i})\cos\varphi_{i},
\sin(u\vartheta_{i})\sin\varphi_{i}, \cos(u\vartheta_{i})) , \eqa
where $u$ is an additional time-like parameter for the Berry phase
term.\cite{Nagaosa} We note that $n_{i}^{+} =
\sin\vartheta_{i}e^{i\varphi_{i}}$ corresponds to the pairing
potential $\Phi_{i} = \Phi_{i}^{R} + i\Phi_{i}^{I}$. Inserting Eq.
(11) into Eq. (10), and performing the integration over $u$ in the
Berry phase term, we obtain the following expression for the
nonlinear $\sigma$ model \bqa && S_{eff} =
iS\sum_{i}(-1)^{i}\int_{0}^{c\beta}{dx_{0}}(1-\cos{\vartheta}_{i})\dot{\varphi}_{i}
\nn && +
\int_{0}^{c\beta}dx_{0}\int{d^d\mathbf{x}}\frac{1}{g}[\sin^{2}\vartheta(\partial_{\mu}\varphi)^{2}
+ (\partial_{\mu}\vartheta)^{2}] \nn && +
\int_{0}^{c\beta}dx_{0}\int{d^d\mathbf{x}}\frac{1}{g}[ -
(\mu+v)^{2}\sin^{2}\vartheta +
4i(\mu+v){\dot\varphi}\sin^{2}\vartheta ] \nn && + S_{I} , \nn &&
S_{I} = I\int_{0}^{c\beta}dx_{0}\int{d^d\mathbf{x}}
\cos^{2}\vartheta , \eqa where we introduced the action $S_{I}$
favoring the XY order. This procedure is quite parallel to that in
the SO(5) $\sigma$ model.\cite{Tanaka} The chemical potential
favors the XY order without the "easy plane" anisotropy term. The
easy plane anisotropy allows us to set $\vartheta_{i} = \pi/2$. In
this case Eq. (12) reads \bqa && S_{XY} = i\pi\sum_{i}[(-1)^{i} +
\frac{8}{g}(\mu+v_{i})]q_{i} \nn && +
\int_{0}^{c\beta}dx_{0}\int{d^d\mathbf{x}}\Bigl[\frac{1}{2u_{\varphi}}\dot{\varphi}^{2}
+ \frac{\rho_{\varphi}}{2}(\nabla_{\mathbf{x}}\varphi)^{2} \Bigr]
. \eqa Here $q_{i} =
(1/2\pi)\int_{0}^{c\beta}{dx_{0}}\dot{\varphi}_{i}$ is an integer
representing an instanton number, here a vortex charge, and the
pseudospin value $S = 1/2$ is used. Anisotropy in time and spatial
fluctuations of the $\varphi$ fields is introduced by
$u_{\varphi}$ and $\rho_{\varphi}$. The effective field theory for
the SC-CDW transition is given by the quantum XY model with Berry
phase in the easy plane limit of Eq. (10). It is clear that the
topological phase appears even at half filling as a result of the
competition between SC and CDW. The chemical potential plays the
role of an additional Berry phase in the phase field $\varphi$.

\subsection{Effective vortex action with both external and random dual magnetic
flux}

To take into account the Berry phase contribution, we resort to a
duality transformation, and obtain the dual vortex action \bqa &&
S_{v} = - t_{v}
\sum_{nm}\Phi_{n}^{\dagger}e^{i\bar{c}_{nm}+ic_{nm}}\Phi_{m} +
V(|\Phi_{n}|) \nn && +
\frac{1}{2e_{v}^{2}}\sum_{\mu}(\partial\times{c})_{\mu}^{2} -
\frac{4}{g{e}_{v}^{2}}\sum_{\mu}v_{i}(\nabla\times{c})_{i} . \eqa
Here $\Phi_{n}$ is a vortex field residing in the $(2+1)D$ dual
lattice $n$ of the original lattice $\mu = (\tau, i)$, and
$c_{nm}$ a vortex gauge field. $V(|\Phi_{n}|)$ is an effective
vortex potential. $e_{v}$ is a coupling constant of the vortex
field to the vortex gauge field. $\bar{c}_{nm}$ is a background
gauge potential for the vortex field, resulting from the Berry
phase contribution and satisfying at half filling \bqa &&
(\nabla\times\bar{c})_{i} = (-1)^{i}\pi . \nonumber \eqa
Randomness $v_{i}$ plays the role of a dual random magnetic field
in vortices.

In the mean field approximation ignoring vortex-gauge fluctuations
$c_{nm}$, one finds that the vortex problem coincides with the
well known Hofstadter one. If one considers a dual magnetic flux
$f = p/q$ with relatively prime integers $p, q$ (here, $p =1$ and
$q = 2$), the dual vortex action has $q$-fold degenerate minima in
the magnetic Brillouin zone. Low energy fluctuations near the
$q$-fold degenerate vacua are assigned to be $\psi_{l}$ with $l =
0, ..., q-1$. Balents et al. constructed an effective LGW free
energy functional in terms of low energy vortex fields $\Psi_{l}$,
given by linear combinations of $\psi_{l}$.\cite{Balents}
Constraints for the effective potential of $\Psi_{l}$ are symmetry
properties associated with lattice translations and rotations in
the presence of the dual magnetic field. In the present $q = 2$
case (corresponding to a $\pi$ flux phase) there are two
degenerate vortex ground states at momentum $(0,0)$ and
$(\pi,\pi)$. Introducing the linear-combined vortex fields of
$\Psi_{0} = \psi_{0} + i\psi_{1}$ and $\Psi_{1} = \psi_{0} -
i\psi_{1}$ where $\psi_{0}$ and $\psi_{1}$ are the low energy
vortex fluctuations around the two degenerate ground states
respectively, and considering the symmetry properties mentioned
above, one can find an effective low energy action. However, one
important difference from the previous study\cite{Balents} due to
the contribution of random Berry phase should be taken into
account carefully. One cautious person may doubt if it is
meaningful to consider the magnetic Brillouin zone in the presence
of randomness. Actually, this is a correct question. In this paper
we assume the existence of the magnetic Brillouin zone since the
limit of weak randomness is of our interest.

Based on symmetry properties of the square lattice under $\pi$
flux, we write down the effective action for low energy vortices
with randomness \bqa && S_{eff} = \int{d\tau}d^2r\Bigl[
|(\partial_{\mu} - ic_{\mu})\Psi_{0}|^{2} + |(\partial_{\mu} -
ic_{\mu})\Psi_{1}|^{2} \nn && + m^{2}(|\Psi_{0}|^{2} +
|\Psi_{1}|^{2}) + u_{4}(|\Psi_{0}|^{2} + |\Psi_{1}|^{2})^{2} \nn
&& + v_{4}|\Psi_{0}|^{2}|\Psi_{1}|^{2} -
v_{2}(\Psi_{0}^{*}\Psi_{1} + H.c.) \nn && +
\frac{1}{2e_{v}^{2}}(\partial\times{c})^{2} \Bigr] -
\int{d\tau}d^2rv(\partial\times{c})_{\tau} . \eqa In the effective
vortex potential $m^{2}$ is a vortex mass, $u_{4}$ a local
interaction, $v_{4}$ a cubic anisotropy, and $v_{2}$ breaking the
U(1) phase transformation $\Psi_{0(1)} \rightarrow
e^{i\varphi_{0(1)}}\Psi_{0(1)}$ in the presence of random Berry
phase for vortices. There are two important differences between
the cases with and without disorder. In the absence of disorder
the $v_{2}$ term is given by $- v_{8}[(\Psi_{0}^{*}\Psi_{1})^{4} +
H.c.]$ owing to the four-fold symmetry.\cite{Balents,Senthil_DQCP}
However, the presence of weak disorder implies that lattice
translations and rotations are no longer symmetries. This reduces
the fourth power to the first one. Furthermore, we estimate that
$v_{2}$ is a random variable depending on disorder. One can regard
$v_{2}$ as an instanton fugacity.\cite{Senthil_DQCP,Kim_DQCP}
Thus, the estimation of the random variable $v_{2}$ means that
disorder makes the instanton fugacity random. As another
contribution of disorder $v$ is a dual random magnetic field in
the last term. This term generates different kinds of random
potentials, as will be seen later.

Based on the effective vortex potential Eq. (15), one can perform
a mean field analysis in the absence of disorder ($v =
0$).\cite{Lannert} Condensation of vortices occurs in the case of
$m^{2} < 0$ and $u_{4} > 0$. The signs of $v_{4}$ and $v_{8}$
determine the ground state. For $v_{4} < 0$, both vortices have a
nonzero vacuum expectation value $|\langle\Psi_{0}\rangle| =
|\langle\Psi_{1}\rangle| \not= 0$, and their relative phase is
determined by the sign of $v_{8}$. In the case of $v_{8} > 0$ the
resulting vortex state corresponds to a columnar dimer order,
breaking both the rotational and translational symmetries. In the
case of $v_{8} < 0$ the resulting phase exhibits a plaquette
pattern, braking the rotational symmetries. On the other hand, if
$v_{4} > 0$, the ground states are given by either
$|\langle\Psi_{0}\rangle| \not= 0, |\langle\Psi_{1}\rangle| = 0$
or $|\langle\Psi_{0}\rangle| = 0, |\langle\Psi_{1}\rangle| \not=
0$, and the sign of $v_{8}$ is irrelevant. In this case an
ordinary charge density wave order at wave vector $(\pi, \pi)$ is
obtained, breaking the translational symmetries. This mean field
analysis coincides with that in Ref. \cite{Senthil_DQCP}.

At the critical point $m^{2} = 0$ the eighth-order term is
certainly irrelevant owing to its high order. Furthermore, the
cubic anisotropy term ($v_{4}$) is well known to be irrelevant in
the case of $q < q_{c} = 4$, ignoring vortex gauge
fluctuations.\cite{Cubic} As a result, the Heisenberg fixed point
($v_{4}^{*} = 0$ and $u_{4}^{*} \not= 0$) appears in the limit of
zero vortex charge ($e_{v} \rightarrow 0$). Allowing the vortex
gauge fields at the Heisenberg fixed point, the Heisenberg fixed
point becomes unstable, and a new fixed point with a nonzero
vortex charge appears as long as the cubic anisotropy $v_{4}$ is
assumed to be irrelevant.\cite{Herbut_BG,Kim_disorder} This
charged fixed point seems to be qualitatively the same as that
obtained in the absence of the dual magnetic field, i.e., the $q =
1$ case. However, one important difference is that the dual flux
quantum (corresponding to an electromagnetic charge of the
original boson) seen by the vortex field $\Psi_{0(1)}$ is halved
due to the two flavors of vortices.\cite{Balents} This implies
that the boson excitations dual to the vortices carry an
electromagnetic charge $e$ instead of $2e$. These fractionalized
excitations are confined to appear as usual Cooper pair
excitations with charge $2e$ away from the quantum critical point,
resulting from the eighth-order term to break the U(1) gauge
symmetry.\cite{Kim_DQCP} However, as mentioned above, this $v_{8}$
term becomes irrelevant at the critical point, indicating that the
charge-fractionalized excitations are deconfined to appear. Thus,
the SC-CDW transition at half filling occurs via the deconfined
quantum critical point as the AF-VBS
transition.\cite{Senthil_DQCP} This conclusion does not depend on
whether the cubic anisotropy is relevant or not at the charged
critical point. Even if $v_{4}$ is relevant at the isotropic
charged fixed point to cause a new anisotropic charged fixed
point, the eighth-order term associated with charge
fractionalization would be irrelevant.

\section{Role of disorder in the deconfined quantum critical point}

Now we investigate the role of disorder in the deconfined quantum
critical point. In order to take into account the random
potentials by disorder, we use the replica trick to average over
disorder. The random magnetic field $v$ and the random fugacity
$v_{2}$ in the vortex action Eq. (15) would cause \bqa && -
\sum_{k,k' = 1}^{N}\int{d\tau}{d\tau_1}\int{d^2r}\frac{\Im}{2}
(\partial\times{c}_{k})_{\tau}(\partial\times{c}_{k'})_{\tau_1} ,
\nn && - \sum_{k,k' =
1}^{N}\int{d\tau}{d\tau_1}\int{d^2r}\frac{\Re}{2}
(\Psi_{0k}^{*}\Psi_{1k} +
H.c.)_{\tau}\nn&&~~~~~~~~~~~~~~~~~~~~~~~~~~~~ \times
(\Psi_{0k'}^{*}\Psi_{1k'} + H.c.)_{\tau_1} \nonumber \eqa for
Gaussian random potentials satisfying \bqa && \langle{v}(r)\rangle
= 0, ~~~~~ \langle{v}(r)v(r_1)\rangle = \Im \delta(r-r_1) , \nn &&
\langle{v}_{2}(r)\rangle = 0, ~~~~~
\langle{v}_{2}(r)v_{2}(r_1)\rangle = \Re \delta(r-r_1) \nonumber
\eqa with the strength $\Im$ and $\Re$ of the random potentials,
respectively. Here $k, k' = 1, ..., N$ denote replica indices, and
the limit $N \rightarrow 0$ is done at the final stage of
calculations. However, inclusion of only this correlation term is
argued to be not enough for disorder effects. Because the
gauge-field propagator has off-diagonal components in replica
indices, the vortex-gauge interaction of the order
$\Im^{2}e_{v}^{4}$ generates a quartic term including the
couplings of different replicas of vortices even if this term is
absent initially.\cite{Herbut_BG} The resulting disordered vortex
action is obtained to be \bqa && Z_{R} =
\int{D\Psi_{0k}D\Psi_{1k}D{c}_{k\mu}} e^{-S_{R}} , \nn && S_{R} =
S_{v} + S_{d} + S_{f} , \nn && S_{v} = \sum_{k =
1}^{N}\int{d\tau}d^2r\Bigl[ |(\partial_{\mu} -
ic_{k\mu})\Psi_{0k}|^{2} + |(\partial_{\mu} -
ic_{k\mu})\Psi_{1k}|^{2} \nn && + m^{2}(|\Psi_{0k}|^{2} +
|\Psi_{1k}|^{2}) + u_{4}(|\Psi_{0k}|^{2} + |\Psi_{1k}|^{2})^{2}
\nn && + v_{4}|\Psi_{0k}|^{2}|\Psi_{1k}|^{2} +
\frac{1}{2e_{v}^{2}}(\partial\times{c}_{k})^{2} \Bigr] , \nn &&
S_{d} = - \sum_{k,k' =
1}^{N}\int{d\tau}{d\tau_1}\int{d^2r}\nn&&\frac{\Re}{2}
(\Psi_{0k}^{*}\Psi_{1k} + H.c.)_{\tau} (\Psi_{0k'}^{*}\Psi_{1k'} +
H.c.)_{\tau_1} \nn && - \sum_{k,k' = 1}^{N}\sum_{q,q' =
0}^{1}\int{d\tau}{d\tau_1}\int{d^2r}\frac{W}{2}|\Psi_{qk\tau}|^{2}|\Psi_{q'k'\tau_1}|^{2}
, \nn && S_{f} = - \sum_{k,k' =
1}^{N}\int{d\tau}{d\tau_1}\int{d^2r}\frac{\Im}{2}(\partial\times{c}_{k})_{\tau}(\partial\times{c}_{k'})_{\tau_1}
\eqa with $W > 0$. The last term induced by disorder in $S_{d}$
has the same form with the term resulting from a random mass term.
The correlation term $S_{f}$ between random magnetic fluxes would
be ignored in this paper. In the small $\Im$ limit this term was
shown to be exactly marginal at one loop level.\cite{Herbut_BG}

The question is what happens on the deconfined charged critical
point when randomness is turned on. It is not an easy task to take
into account all of the terms on an equal footing in the RG
analysis. To investigate the role of the two disorder-induced
terms of $S_{d}$ in the deconfined charged critical point, one can
consider two approximate ways. One is first to examine the random
mass term, denoted by the coupling strength $W$, at the deconfined
charged critical point, and then to see what happens if the random
fugacity ($\Re$) is turned on at a weak disorder fixed point. The
other is first to investigate the effect of the random fugacity
term on the deconfined charged critical point, and then to examine
the random mass term. In this paper we follow the second approach
because our main interest is to see the fate of the deconfined
quantum criticality against randomness. It should be noted that
the existence of the deconfined quantum criticality is determined
by the fugacity term.\cite{Kim_DQCP}

To examine the role of the random fugacity term in the charged
critical point, we consider a phase-only action ignoring amplitude
fluctuations of vortices.\cite{London} This so-called London
approximation was also utilized in Refs.
\cite{Senthil_DQCP,Kim_DQCP,Balents}. The effective vortex action
is obtained to be \bqa && S_{R} = \sum_{k =
1}^{N}\int{d\tau}d^2r\Bigl[\sum_{q=0}^{1}\frac{\rho}{2}
(\partial_{\mu}\theta_{qk} - c_{k\mu})^{2} +
\frac{1}{2e_{v}^{2}}(\partial\times{c}_{k})^{2} \Bigr] \nn && -
\sum_{k,k' = 1}^{N}\int{d\tau}{d\tau_1}\int{d^2r}
\frac{\Re}{2}\cos(\theta_{0k}-\theta_{1k})_{\tau}
\cos(\theta_{0k'}-\theta_{1k'})_{\tau_1} , \nn \eqa where $\rho$
is a stiffness parameter proportional to the condensation
probability of vortices in the mean field level. The parameter
$\Re$ is also renormalized by the condensation amplitude of
vortices.

To see whether the random $\cos$ term is relevant or not at the
charged fixed point, it is necessary to check the existence of the
charged critical point without the disorder-induced term.
Considering $\Re = 0$ in Eq. (17), we obtain the RG equations for
the stiffness $\rho$ and the vortex charge $e_{v}^{2}$ \bqa &&
\frac{d\rho}{dl} = \rho - \gamma{e_{v}^{2}}{\rho} , \nn &&
\frac{de_{v}^{2}}{dl} = e_{v}^{2} - 2\lambda{e_{v}^4} , \eqa where
$\gamma$ and $\lambda$ are positive numerical
constants,\cite{Charged_fixed_point} and $l$ is a usual scaling
parameter. The last term $- \gamma{e_{v}^2}\rho$ in the first
equation originates from the self-energy correction of the vortex
field owing to gauge fluctuations while the term $-
\lambda{e}_{v}^4$ in the second equation results from that of the
gauge field due to screening of the vortex charge. In these RG
equations there exist two fixed points; one is the neutral (XY)
fixed point of $e_{v}^{*2} = 0$ and $\rho^{*} = 0$ and the other,
the charged (IXY) fixed point of $e_{v}^{*2} = \frac{1}{\lambda}$
and $\rho^{*} = 0$. The neutral fixed point is unstable against a
nonzero charge $e_{v}^2 \not= 0$, and the RG flows in the
parameter space of $(\rho,e_{v}^{2})$ converge into the charged
fixed point owing to $1 - \gamma{e_{v}^{*2}} = 1 -
\frac{\gamma}{\lambda} < 0$.\cite{Kim_DQCP}

Next we examine the role of the random fugacity term ignoring
vortex gauge fluctuations, i.e., $e_{v}^{2} = 0$. The random
fugacity term can be rewritten in the following way \bqa &&
\frac{\Re}{2}\cos(\theta_{0k}-\theta_{1k})_{\tau}
\cos(\theta_{0k'}-\theta_{1k'})_{\tau_1} \nn && =
\frac{\Re}{4}\cos[(\theta_{0k}-\theta_{1k})_{\tau} +
(\theta_{0k'}-\theta_{1k'})_{\tau_1}] \nn && +
\frac{\Re}{4}\cos[(\theta_{0k}-\theta_{1k})_{\tau} -
(\theta_{0k'}-\theta_{1k'})_{\tau_1}] . \eqa In this expression we
can find that the last term is the most relevant term owing to its
sign. Thus, it is reasonable to consider the following action for
the RG analysis \bqa && S_{R} \approx \sum_{k =
1}^{N}\int{d\tau}d^2r \Bigl[ \frac{\rho}{2}
(\partial_{\mu}\theta_{0k})^{2} + \frac{\rho}{2}
(\partial_{\mu}\theta_{1k})^{2} \Bigr] \nn && - \sum_{k,k' =
1}^{N}\int{d\tau}{d\tau_1}\int{d^2r}
\frac{\Re}{4}\cos[(\theta_{0k}-\theta_{1k})_{\tau}
-(\theta_{0k'}-\theta_{1k'})_{\tau_1}] . \nonumber \eqa This
action was well studied in the context of Anderson localization in
one dimensional systems when the flavor number of bosons is
one.\cite{One_dimension} In Ref. \cite{Kim_DQCP} we derived RG
equations for the two-flavor sine-Gordon action. Similarly, one
can easily obtain the following RG equations for the stiffness
$\rho$ and the random parameter $\Re$ \bqa && \frac{d\rho}{dl} =
\rho + {\beta}{\Re^2} \frac{2}{\rho} , \nn && \frac{d\Re}{dl} = (4
- \alpha\frac{2}{\rho})\Re \eqa with positive numerical constants,
$\beta$ and $\alpha$. In our consideration their precise values
are not important. The effect of two flavors appears as the factor
$2$ in the $1/\rho$ terms. One important difference between the
present $(2+1)D$ study and the previous $(1+1)D$
one\cite{One_dimension} is that the bare scaling dimensions of
$\rho$ and $\Re$ are given by $1$ and $4$ in $(2+1)D$ while $0$
and $3$ in $(1+1)D$, respectively. This difference results in the
fact that there exist no stable fixed points in $(2+1)D$ while in
$(1+1)D$ there is a line of fixed points describing the
Kosterliz-Thouless transition.\cite{Herbut_BG,One_dimension} Both
the phase stiffness $\rho$ and the parameter $\Re$ become larger
and larger at low energy. This implies that depth of the random
$\cos$ potential in Eq. (17) becomes deeper and deeper, making the
phase difference $\theta_{0} - \theta_{1}$ pinned at one ground
position of the $\cos$ potential. This is the signal of
confinement between fractionalized excitations, $\theta_{0}$ and
$\theta_{1}$.\cite{Kim_DQCP}

Combining Eq. (18) with Eq. (20), we obtain the RG equations for
the stiffness $\rho$, the vortex charge $e_{v}^{2}$, and the
random parameter $\Re$ \bqa && \frac{d\rho}{dl} = \rho -
\gamma{e_{v}^{2}}{\rho} + {\beta}{\Re^2} \frac{2}{\rho}  , \nn &&
\frac{de_{v}^{2}}{dl} = e_{v}^{2} - 2\lambda{e_{v}^{4}} , \nn &&
\frac{d\Re}{dl} = (4 - \alpha\frac{2}{\rho})\Re . \eqa These RG
equations tell us that the nonzero fixed point value of the vortex
charge ($e_{v}^{2*} = \frac{1}{2\lambda}$) in the second RG
equation makes the stiffness parameter $\rho$ vanish ($\rho^* =
0$) in the first RG equation, causing the random parameter to be
irrelevant, i.e., $\Re^{*} = 0$ in the third RG equation. This
solution is self-consistent with the first RG equation. This
result means that as long as the stable charged fixed point
exists, the random fugacity term is irrelevant at the charged
critical point. As a result, we find only one stable fixed point
of $e_{v}^{2*} = \frac{1}{2\lambda}$, $\rho^* = 0$ and $\Re^{*} =
0$. The deconfined quantum criticality is stable against the
random fugacity term.

Now we consider the random mass term at this deconfined charged
critical point. At the tree level one can easily check that the
random mass term is relevant at the charged critical point,
indicating instability of the charged fixed point against
disorder. One-loop RG analysis shows that a weak disorder fixed
point appears if the cubic anisotropy is
irrelevant.\cite{Herbut_BG,Kim_disorder} One important point is
that the fixed point value of a vortex charge is nonzero at the
weak disorder fixed point, given by the value $e_{v}^{2*} =
\frac{1}{2\lambda}$ of the charged critical
point.\cite{Herbut_BG,Kim_disorder} Furthermore, the fixed point
value of the phase stiffness would still be zero at the random
charged critical point because the vortex condensation should
occur at $\rho^{*} = 0$. Based on this discussion, we expect that
the random fugacity term would still be zero at the weak disorder
fixed point. This implies that although the dimerised or CDW
phases may be unstable owing to disorder, turning into glassy
phases, deconfinement of fractionalized charge excitations is
expected to survive at the disorder critical point. However, we
admit that because we did not treat the two disorder-induced terms
of $S_{d}$ in Eq. (16) on an equal footing, the present result is
not fully justified.

\section{Summary and Discussion}

In summary, we showed that the competition between
superconductivity (SC) and charge density waves (CDWs) results in
a non-trivial Berry phase for the SC and CDW order parameters even
at half filling, allowing a deconfined quantum critical point of
fractionalized charge excitations with $e$ instead of $2e$. We
considered the stability of the deconfined quantum criticality
against quenched randomness generating two kinds of random
potentials, a random mass term and a random fugacity one in the
vortex action. Within the London approximation we showed that the
random fugacity term is irrelevant at the charged critical point.
Then, we discussed the effect of the random mass term on this
fixed point, and found that the charged critical point becomes
unstable, and a weak disorder fixed point with a nonzero vortex
charge appears. We argued that since the random fugacity term
would still be irrelevant at this disorder fixed point owing to
the finite fixed point value of the vortex charge, deconfinement
of fractionalized excitations survives in the weak disorder limit.

A cautious person may ask the relevance of this LGW-forbidden
quantum transition because there has been no clear indication in
actual physical systems so far. One way to justify this quantum
transition is to find its one dimensional analogue. Considering
spin fluctuations associated with the AF-VBS transition, its
critical field theory is well known to be an effective O(4)
nonlinear $\sigma$ model with a topological $\theta$ term as an
SU(2) level-1 Wess-Zumino-Witten (WZW)
theory.\cite{Senthil_superspin} This effective field theory can be
derived from some microscopic models such as the bond-alternating
spin chain\cite{AF_VBS_1D} and the Peierls-Hubbard
model\cite{Tanaka_PHM} via non-Abelian bosonization. We believe
that this procedure can be applied to charge fluctuations
associated with competition between SC and CDWs. Actually, Carr
and Tsvelik investigated the continuous SC-CDW transition in a
quasi-one-dimensional system.\cite{Tsvelik} They considered an
effective model of spin-gapped chains weakly coupled by Josephson
and Coulomb interactions. They obtained an effective field theory
for SC and CDW fluctuations in the framework of the non-Abelian
bosonization with weak interchain-interactions. They found its
phase diagram to show the SC and CDW phases, separated by line of
critical points which exhibits an approximate SU(2) (charge)
symmetry. They proposed that the critical line would shrink to a
point in two dimensions, identified with the quantum critical
point in the SC-CDW quantum transition. Furthermore, they
discussed the relevance of their theory, considering the
experimental system of $Sr_2Ca_{12}Cu_{24}O_{41}$ built up from
alternating layers of weakly coupled $CuO_{2}$ chains and
$Cu_{2}O_{3}$ two-leg ladders. One important difference is that
the effective field theory in Ref. \cite{Tsvelik} does not include
a topological $\theta$ term while our field theory does allow the
$\theta$ term. In this respect the correspondence between the
present description and the previous theory\cite{Tsvelik} is not
complete. A further investigation for the one-dimensional system
is necessary near future.

An important future work in this direction is to introduce spin
degrees of freedom associated with an antiferromagnetic order.
Then, the resulting effective nonlinear $\sigma$ model would
posses an SO(4)$\cong$SU(2)$\bigotimes$SU(2) symmetry, where the
former SU(2) is associated with spin, and the latter SU(2)
pseudospin. A topological term would appear in this SO(4) $\sigma$
model. The competition between antiferromagnetism,
superconductivity, and density waves remains to be solved.

K.-S. Kim would like to thank Dr. A. Tanaka for his kind
explanation of the conflict in Refs. \cite{Oshikawa,DHLee}.


\begin{thebibliography}{9}
\bibitem{Oshikawa} M. Oshikawa, Phys. Rev. Lett. {\bf 84}, 1535
(2000).
\bibitem{DHLee} D. H. Lee and R. Shankar, Phys. Rev. Lett. {\bf
65}, 1490 (1990).
\bibitem{Tanaka} A. Tanaka and X. Hu, Phys. Rev. Lett.
{\bf 95}, 036402 (2005).
\bibitem{Senthil_superspin} T. Senthil and M. P. A. Fisher,
Phys. Rev. B {\bf 74}, 064405 (2006).
\bibitem{Senthil_DQCP} T. Senthil, A. Vishwanath, L. Balents, S.
Sachdev, and M. P. A. Fisher, Science {\bf 303}, 1490 (2004); T.
Senthil, L. Balents, S. Sachdev, A. Vishwanath, and M. P.A.
Fisher, Phys. Rev. B {\bf 70}, 144407 (2004).
\bibitem{Kim_DQCP} Ki-Seok Kim, Phys. Rev. B {\bf 72}, 035109
(2005).
\bibitem{Old_SC_CDW} K. Borejsza and N. Dupuis, Phys. Rev. B
{\bf 69}, 085119 (2004); C. N. Yang, Phys. Rev. Lett. {\bf 63},
2144 (1989); C. N. Yang and S. C. Zhang, Mod. Phys. Lett. B {\bf
4}, 759 (1990); I. F. Herbut, Phys. Rev. B {\bf 60}, 14503 (1999).
\bibitem{Hard_core_boson} M. Keller, W. Metzner, and U.
Schollwock, Phys. Rev. Lett. {\bf 86}, 4612 (2001), and references
therein.
\bibitem{Zhang} E. Demler, W. Hanke, and S.-C. Zhang,
Rev. Mod. Phys. {\bf 76}, 909 (2004).
\bibitem{Balents} L. Balents, L. Bartosch, A. Burkov, S. Sachdev,
and K. Sengupta, Phys. Rev. B {\bf 71}, 144508 (2005).
\bibitem{Tesanovic} Z. Tesanovic, Phys. Rev. Lett. {\bf 93},
217004 (2004).
\bibitem{Kim_Kondo} K.-S. Kim, Phys. Rev. B {\bf 72}, 144426
(2005).
\bibitem{Wen} X. G. Wen and A. Zee, Phys. Rev. Lett. {\bf 61},
1025 (1988).
\bibitem{Nagaosa} N. Nagaosa, Quantum Field Theory in
Strongly Correlated Electronic Systems, Springer (1999).
\bibitem{Lannert} C. Lannert, M. P. A. Fisher, and T.
Senthil, Phys. Rev. B {\bf 63}, 134510 (2001).
\bibitem{Cubic} P. M. Chaikin and T. C. Lubensky, Principles of
condensed matter physics, Cambridge University Press (1995), Chap.
5; J. M. Carmona, A. Pelissetto, and E. Vicari, Phys. Rev. B {\bf
61}, 15136 (2000).
\bibitem{Herbut_BG} I. F. Herbut, Phys. Rev. Lett. {\bf 79}, 3502
(1997); I. F. Herbut, Phys. Rev. B {\bf 57}, 13729 (1998).
\bibitem{Kim_disorder} Ki-Seok Kim, Phys. Rev. B {\bf 73}, 235115
(2006); The present author also investigated the stability of
algebraic spin liquid in the weak disorder limit, Ki-Seok Kim,
Phys. Rev. B {\bf 70}, 140405(R) (2004), and Phys. Rev. B {\bf
72}, 014406 (2005).
\bibitem{London} We note that a vortex action is given by
a phase-only action in the duality transformation originally. See
Refs. \cite{Balents,Senthil_DQCP}.
\bibitem{Charged_fixed_point} From the
relation of $\rho_{R} = |\Psi_{R}|^2 = Z^{-1}_{\Psi}|\Psi_{B}|^2 =
Z^{-1}_{\Psi}\rho_{B}$ it is necessary to know the wave function
renormalization constant $Z_{\Psi}$. Here $R$ and $B$ represent
$renormalized$ and $bare$, respectively. The renormalization
factor $Z_{\Psi}$ can be easily obtained from the one-loop
self-energy calculation for the vortex field. The self-energy
$\Sigma(p)$ of the vortex field is given by $\Sigma(p) =
e_{v}^2\int\frac{d^3k}{(2\pi)^3}\frac{1}{|p-k|^2}(2p-k)_{\mu}D_{\mu\nu}(k)(2p-k)_{\nu}$,
where $D_{\mu\nu}(k) = \frac{1}{k^2}\Bigl(\delta_{\mu\nu} -
\frac{k_{\mu}k_{\nu}}{k^2} \Bigr)$ is the propagator of vortex
gauge fields in the Landau gauge. We find $Z_{\Psi}^{-1} = 1 -
\gamma{e}_{v}^2$, where $\gamma$ is a positive numerical constant.
In the same way we can obtain the $RG$ equation for the vortex
charge $e_{v}^2$. From the relation of $e_{R}^{2} =
Z_{c}e_{B}^{2}$, we find the renormalization factor $Z_{c}$ of the
U(1) gauge field $c_{\mu}$. It can be derived from the
polarization function $\Pi_{\mu\nu}(q)$, given by $\Pi_{\mu\nu}(q)
= e_{v}^2\int\frac{d^3k}{(2\pi)^3}
\frac{(2q-k)_{\mu}(2q-k)_{\nu}}{|q-k|^2|k|^2}$. We obtain $Z_{c} =
1 - 2\lambda{e}_{v}^2$, where $\lambda$ is a positive numerical
constant, and the prefactor $2$ in the $e_{v}^{2}$ term results
from the two flavors.
\bibitem{One_dimension} T. Giamarchi and H. J. Schulz, Phys. Rev.
B {\bf 37}, 325 (1988); M. P. A. Fisher, P. B. Weichman, G.
Grinstein, and D. S. Fisher, Phys. Rev. B {\bf 40}, 546 (1989).
\bibitem{AF_VBS_1D} F. D. M. Haldane, J. Appl. Phys. {\bf 57},
3359 (1985); I. Affleck, Nucl. Phys. B {\bf 265}, 409 (1985).
\bibitem{Tanaka_PHM} A. Tanaka and X. Hu, Phys. Rev. Lett. {\bf
88}, 127004 (2002).
\bibitem{Tsvelik} Sam T. Carr and A. M. Tsvelik, Phys. Rev. B {\bf
65}, 195121 (2002).
\end{thebibliography}
\end{document}